\documentclass[prl,showpacs,amsmath,amssymb,amsfonts,lengthcheck,longbibliography,superscriptaddress]{revtex4-2}
\usepackage{graphicx}  % needed for figures
\usepackage{dcolumn}   % needed for some tables
\usepackage{bm}        % for math
\usepackage{braket}
\usepackage[colorlinks=true,citecolor=blue,linkcolor=blue,urlcolor=blue]{hyperref}

\newcommand{\tr}[1]{\,\text{tr}\left\{#1\right\}}

\begin{document}

%\title{Time and Energy Requirements for Quantum Synchronization}
\title{Energetic cost for speedy synchronization in non-Hermitian quantum dynamics}

\author{Maxwell Aifer}
\email{maifer1@umbc.edu}
\affiliation{Department of Physics, University of Maryland, Baltimore County, Baltimore, MD 21250, USA}

\author{Juzar Thingna}
\email{juzar$_$thingna@uml.edu} 
\affiliation{Department of Physics and Applied Physics, University of Massachusetts, Lowell, MA 01854, USA}
\affiliation{Center for Theoretical Physics of Complex Systems, Institute for Basic Science (IBS), Daejeon 34126, Republic of Korea}

\author{Sebastian Deffner}
\email{deffner@umbc.edu}
\thanks{\\ JT and SD contributed equally to this work}
\affiliation{Department of Physics, University of Maryland, Baltimore County, Baltimore, MD 21250, USA}

\date{\today}

\begin{abstract}
Quantum synchronization is crucial for understanding complex dynamics and holds potential applications in quantum computing and communication. Therefore, assessing the thermodynamic resources required for finite-time synchronization in continuous-variable systems is a critical challenge. In the present work, we find these resources to be extensive for large systems. We also bound the speed of quantum and classical synchronization in coupled damped oscillators with non-Hermitian anti-$\mathcal{PT}$-symmetric interactions, and show that the speed of synchronization is limited by the interaction strength relative to the damping. Compared to the classical limit, we find that quantum synchronization is slowed by the non-commutativity of the Hermitian and anti-Hermitian terms. Our general results could be tested experimentally and we suggest an implementation in photonic systems. 
\end{abstract}

\maketitle
 The study of synchronization dates at least to the 17th century, when Huygens noted the gradual build-up of correlations in the motion of coupled pendula \cite{oliveira_huygens_2015}. Similar behavior has since been found ubiquitously in nature, such as in many-body physics, biology, and even human activities \cite{strogatz_spontaneous_1997, strogatz_synchronization_2003, khasseh_many-body_2019, ha_collective_2016, mirollo_synchronization_1990, glass_synchronization_2001, giuricin_synchronization_1984, toiya_synchronization_2010,Denisov20}. As the coordination of multiple objects implies some kind of communication, synchronization is a signature of information flow, and is a key mechanism for establishing order from disorder in complex systems \cite{bollt_synchronization_2012, pethel_information_2003, parrondo_thermodynamics_2015,blekhman_problem_1964, luo_theory_2009, ott_chaos_2002}. As such, it is of interest in thermodynamics, where information is a central quantity~\cite{Deffner2013PRX,barato_cost_2020, manzano_thermodynamics_2018, buca_algebraic_2022, jaseem_quantum_2020, solanki_role_2022, lazarides_periodic_2014}. That synchronization can be seen at the smallest observable scales makes it relevant also in quantum information theory, where it has become an emerging research focus due to potential applications in quantum computing and communication \cite{agnesi_simple_2020, calderaro_fast_2020}.

In quantum dynamics, the primary focus has been on synchronization in discrete systems \cite{zhirov_quantum_2009, lohe_quantum_2010, roulet_synchronizing_2018, roulet_quantum_2018, koppenhofer_quantum_2020, TaufiqPRA2023, TaufiqPRL2023}, whereas continuous-variable models are often treated classically~\cite{Ryu15, Herpich18, Ryu_2021}. However, to study the quantum limit of classical models, genuine continuous-variable scenarios are required \cite{Lee13, mari_measures_2013, vacchini_transient_2019, jaseem_generalized_2020, ha_collective_2016, Wachtler23}. Multiple ways of quantifying synchronization have been devised for both discrete and continuous-variable quantum systems \cite{mari_measures_2013, jaseem_generalized_2020,eshaqi-sani_synchronization_2020, hush_spin_2015, li_quantum_2016, li_properties_2017, ameri_mutual_2015}, however there is no clear consensus as to which metric is universally applicable. Moreover, existing work also provides only limited insight into the time and energy scales on which the process occurs.

Most of our current understanding of quantum synchronization is limited to the long-time, steady-state behavior. While this simplifies the description, there are some questions which can only be answered using a finite-time approach, in particular regarding the required resources for synchronization. To this end, some progress has been made using quantum speed limits \cite{deffner_quantum_2017,busch_energy-time_1990-1, busch_energy-time_1990,  deffner_quantum_2020, aifer_quantum_2022, pfeifer_generalized_1995,Poggi2021PRXQ,Fogarty2020PRL,Deffner2013PRL,Deffner2017NJP}, and which are now understood to also constrain the rate of synchronization \cite{impens_shortcut_2023}. Similar results have been proposed using the Lieb-Robinson bound \cite{schmolke_noise-induced_2022}. Yet, the finite-time analysis of synchronization is still in an early developmental stage.

In this letter, we apply quantum speed limits and quantum thermodynamics to a general model of continuous-variable quantum systems. A measure of complete synchronization in continuous-variable systems is defined, which is scale-invariant and is sufficient for phase synchronization. We obtain bounds which relate the degree of synchronization to the distance from thermodynamic equilibrium, resulting in an extensive expression for the minimal work necessary to achieve synchronization.

Specifically, we study a quantum master equation that includes both non-Hermitian dynamics and a dissipative term in the Gorini-Kassakowski-Sudarshan-Lindblad (GKSL) form, resulting in a non-linear dynamical semigroup. We find that the rate of synchronization is determined by a competition between the irreversible entropy production caused by damping, which slows synchronization, and the strength of the anti-Hermitian coupling, which speeds up synchronization. The resulting upper bound on the synchronization rate has terms of the form of the Mandelstam-Tamm inequality \cite{tamm_uncertainty_1991}, where speed scales with the uncertainty of the energy, except in this case even the uncertainties of the Hermitian and anti-Hermitian parts of the Hamiltonian are crucial. As an example, we consider a dissipatively coupled photonic dimer. The classical counterpart of all our results are obtained and in case of the dimer we find that the model reduces to the celebrated coupled Stuart-Landau oscillators~\cite{Landau44, stuart_1960}. For the dimer model we find that the quantum system synchronizes in a parameter regime wherein it is impossible for the classical model to synchronize, thereby displaying a quantum advantage.

\paragraph{Measure of synchronization}

We consider $N$ quantum oscillators with annihilation operators $\hat{a}_1 \dots \hat{a}_N$. The corresponding dimensionless quadrature operators $\hat{\bold{r}} = (\hat{x}_1, \hat{p}_1, \dots \hat{x}_N, \hat{p}_N)^T$ read~\cite{serafini_quantum_2023,serafini_symplectic_2004, adesso_continuous_2014}
\begin{equation}
    \label{phase-space-coords}
    \hat{x}_j = \frac{\hat{a}_j+\hat{a}_j^\dag}{\sqrt{2}}, \: \:    \hat{p}_j = \frac{\hat{a}_j-\hat{a}_j^\dag}{i\sqrt{2}}.
\end{equation}
For the system to synchronize, the phase space coordinates of the different oscillators need to converge. There can be two distinct types of synchronization: (i) complete synchronization (amplitude and phase synchronization), where the phase space trajectories of multiple subsystems converge, and (ii) phase synchronization, for which tshe phase angles of multiple subsystems converge \cite{ott_chaos_2002}. An intuitive measure to characterize complete synchronization, but not phase synchronization, of a quantum bipartite system is $\mathcal{S}_c = 2 \braket{(\hat{\bold{r}}_2-\hat{\bold{r}}_1)^2}^{-1}$~\cite{mari_measures_2013}. The growth of $\mathcal{S}_c$ does not require phase synchronization,e.g., in case of amplitude death $\mathcal{S}_c$ always grows but there is no phase synchronization~\cite{vacchini_transient_2019}. Therefore, we define a new measure of \emph{complete} synchronization by looking at the distance between the oscillators \emph{relative} to the total radius in phase space. For a bipartite system, we have
\begin{equation}
\label{bipartite-D-def}
D^2 \equiv \frac{\braket{(\hat{\bold{r}}_2 - \hat{\bold{r}}_1)^2}}{\braket{\hat{\bold{r}}^2}},
\end{equation}
and we note that the so-defined $D$ is scale-invariant with respect to $\hat{\bold{r}}$. Hence, the degree of synchronization does not change when the system is viewed at different magnifications. The bipartite distance measure can be expressed in terms of angular and radial measures of similarity,
\begin{equation}
\label{bipartite-D-def-2}
    D^2 = 1 - \mathcal{S}_r \mathcal{S}_\theta,
\end{equation}
where 
\begin{equation}
\label{angle-similarity-measure}
    \mathcal{S}_\theta = \frac{\braket{\hat{\bold{r}}_1\cdot \hat{\bold{r}}_2}}{\left|\braket{\hat{\bold{r}}_1}\right| \left|\braket{\hat{\bold{r}}_2} \right|} \equiv \cos \theta,
\end{equation}
and
\begin{equation}
    \label{radial-similarity-measure}
    \mathcal{S}_r = 2\sqrt{\frac{\braket{\hat{\bold{r}}_1^2}}{\braket{\hat{\bold{r}}^2}}\left( 1- \frac{\braket{\hat{\bold{r}}_1^2}}{\braket{\hat{\bold{r}}^2}}\right)}.
\end{equation}
The quantity $\mathcal{S}_r$ defined above is similar to the binary entropy function \cite{mackay_information_2003}, and is maximized when $\braket{\hat{\bold{r}}_1^2} = \braket{\hat{\bold{r}}_2^2}$ where $\mathcal{S}_r=1$. Equations~\eqref{angle-similarity-measure} and \eqref{radial-similarity-measure} reveal that $D^2$ is between $0$ and $2$, with values less than $1$ indicating synchronization and values greater than $1$ indicating anti-synchronization. It is also clear from the form of Eq.~\eqref{bipartite-D-def-2} that for $D^2$ to become small, both $\mathcal{S}_r$ and $\mathcal{S}_\theta$ must approach their maximal values of $1$, implying that we capture complete and phase synchronization by requiring a decay of $D^2$ in time. For a system of $N$ oscillators, Eq.~\eqref{bipartite-D-def} can be generalized as
\begin{equation}
\label{N-mode-D-def}
    D^2 \equiv 2 \left(1 - \frac{\braket{\bar{\bold{r}}^2}}{\braket{\overline{\bold{r}^2}}}\right),
\end{equation}
where $\bar{\bold{r}} = \left( \sum_{j=1}^N \hat{x}_j,\sum_{j=1}^N \hat{p}_j \right)^T/N$ and $\braket{\overline{\bold{r}^2}} = \braket{\bold{r}^2}/N$. Equation~\eqref{N-mode-D-def} reduces to Eq.~\eqref{bipartite-D-def} for two oscillators, and intuitively captures the notion of synchronization in phase space. Moreover, $D^2$ is non-negative, which follows from Jensen's inequality.

Throughout this work, we consider scenarios in which the $N$ oscillators are initially uncoupled and separately in contact with a thermal bath at inverse temperature $\beta$. Initially, the oscillators are allowed to come to their respective equilibrium states $\hat{\rho}_j^\text{eq}$, and then a coupling between them is turned on. Hence, $\hat{\rho}_0 = \bigotimes_{j=1}^N\hat{\rho}_j^\text{eq}$ is our initial state. To quantify if our system has synchronized we will require that the distance $D$ becomes small and then \emph{stays} small. In other words, given a $D_s$, the system synchronizes to within the distance $D_s$ if there exists a time $\tau$ such that for all $t\geq \tau$, $D\leq D_s$. The smallest $\tau$ for which this holds will be called the synchronization time $\tau_s$.

\paragraph{Dynamics}

Further, the considered $N$ oscillators have natural frequencies $\omega_1, \cdots, \omega_N$ and an arbitrary anti-Hermitian coupling. The corresponding Hamiltonian reads
\begin{equation}
\label{H-def}
\hat{H} = \hat{H}_0 + i \hat{H}_c =  \sum_{j=1}^N  \omega_j \left(\hat{n}_j + \frac{1}{2}\right)  + i \hat{H}_c,
\end{equation}
where $\hat{H}_c$ is Hermitian, $\hat{n}_j$ is the number operator $\hat{a}_j^\dag \hat{a}_j$, and we set $\hbar = 1$. Non-Hermitian Hamiltonians such as the one in Eq.~\eqref{H-def} are effective descriptions of \emph{controlled} dissipation in open quantum systems \cite{ashida_non-hermitian_2020, breuer_theory_2002, bender_real_1998, song_non-hermitian_2019, rotter_non-hermitian_2009, ju_non-hermitian_2019, moiseyev_non-hermitian_2011, Gundogdu_20}. In addition, our system interacts with a thermal environment~\cite{Thingna12,Becker22}, yielding the following quantum master equation (QME)
\begin{equation}
\label{full-QME}
\frac{d\hat{\rho}}{dt} = -i[\hat{H}_0, \hat{\rho}] + \{\hat{H}_c, \hat{\rho}\} - 2\braket{\hat{H}_c}\hat{\rho} + \mathcal{D}[\hat{\rho}],
\end{equation}
where $\braket{\hat{O}} = \text{tr}\{\hat{O} \hat{\rho}\}$, and the term $-2\braket{\hat{H}_c} \hat{\rho}$ is included to preserve normalization. This \emph{nonlinear} equation satisfies the convex quasi-linearity condition and the semigroup property making it a valid quantum evolution~\cite{rembielinski_nonlinear_2021}. The dissipator $\mathcal{D}$ takes the GKSL form     $\mathcal{D}[\hat{\rho}]=\sum_i \hat{F}_i \hat{\rho} \hat{F}_i^\dag - \left\{ \hat{F}_i^\dag \hat{F}_i,\hat{\rho}\right\}/2$. We also split the Liouvillian into non-interacting $\mathcal{L}_0[\hat{\rho}]$ and interacting $\mathcal{L}_c[\hat{\rho}]$ parts given by,
\begin{equation}
\begin{split}
    \mathcal{L}_0[\hat{\rho}] &= -i[\hat{H}_0,\hat{\rho}]+\mathcal{D}[\hat{\rho}] \\  
    \mathcal{L}_c[\hat{\rho}] &= \{\hat{H}_c,\hat{\rho}\}- 2\braket{\hat{H}_c}\hat{\rho},
\end{split}
\end{equation}
such that $d{\hat{\rho}}/dt = \mathcal{L}(\hat{\rho})=\mathcal{L}_0(\hat{\rho})+\mathcal{L}_c(\hat{\rho})$. The state $\hat{\rho}_0$ is a stationary state of the Hermitian $\hat{H}_0$ and Lindblad terms,i.e., $ \mathcal{L}_0(\hat{\rho}_0)=0 $. Note that Eq.~\eqref{full-QME} is nonlinear in $\hat{\rho}$ only because of the term $-2\braket{\hat{H}_c} \hat{\rho}$, as $\braket{\hat{H}_c}$ itself depends linearly on $\hat{\rho}$. We will first examine the case of general $\hat{H}_c$, and later specialize to a specific dimer model which results in a \emph{quantum} Stuart-Landau equation.

\paragraph{Quantum synchronization far from equilibrium}

As the system evolves, it will depart from the initial state $\hat{\rho}_0$ due to the anti-Hermitian coupling. Let $\hat{\rho}_{G;E}$ denote a Gibbs state of the uncoupled system with energy $E$, $\hat{\rho}_{G;E} \propto \exp(-\beta_E \hat{H}_0)$, and $\text{tr}\{\hat{\rho}_{G;E} \hat{H}_0\}=E$. We introduce a measure $\chi$, an \emph{ergotropy}~\cite{Allahverdyan_2004,Sone2012Entropy} \emph{of synchronization}, to quantify the departure of the reduced state $\hat{\rho}$ from the set of Gibbs states of the uncoupled system,
\begin{equation}
\label{chi-def-1}
\chi\equiv\min_U S(\hat{\rho} \| \hat{\rho}_{G;U}),
\end{equation}
in terms of the quantum relative entropy $S(\hat{\rho}_1\|\hat{\rho}_2) = \text{tr}\{\hat{\rho}_1 (\ln \hat{\rho}_1 - \ln \hat{\rho}_2)\}$~\cite{vedral_role_2002}. The minimization over $U$ in Eq.~\eqref{chi-def-1} means that $\chi$ is a property of the state $\hat{\rho}$ and does not depend on the environment parameters (i.e., the bath temperature). This is preferable as the degree of synchronization itself should be independent of the bath parameters, so $\chi$ should be as well to obtain a meaningful relation between them. In the supplemental material~\cite{noauthor_see_nodate}, we show that for the distance measure $D$ to become small, $\chi$ must become large, and we generally have
\begin{equation}
\label{N-mode-quantum-chi-D-bound}
\chi \geq  -2(N-1)\ln \left(\frac{1}{2}e^{1/(N-1)}\sqrt{\frac{(N/\kappa)^{N/(N-1)}}{2(N-1)}}D\right),
\end{equation}
where $\kappa = \omega_\text{min}/\omega_\text{max}$, and for $N=2$ the above general expression reduces to
\begin{equation}
\label{bipartite-quantum-chi-D-bound}
\chi \geq - 2 \ln\left(\frac{eD}{\sqrt{2}\kappa} \right).
\end{equation}

The analogous bound for classical bipartite systems (see supplemental material~\cite{noauthor_see_nodate}) reads
\begin{equation}
\label{bipartite-classical-chi-D-bound}
\chi^\text{(cl)} \geq - 2 \ln\left(\frac{\sqrt{2}D}{\kappa} \right),
\end{equation} 
which is defined in terms of the classical relative entropy \cite{kullback_information_1951, cover_elements_2006}. We note that the results \eqref{bipartite-quantum-chi-D-bound} and \eqref{bipartite-classical-chi-D-bound} have been proved for the case of coupled oscillators, but the proof only relies on certain mild assumptions including the positivity of heat capacity, which is generally valid with some exceptions \cite{schmidt_negative_2001}.

\begin{figure}
\centering
    \includegraphics[width=.48\textwidth]{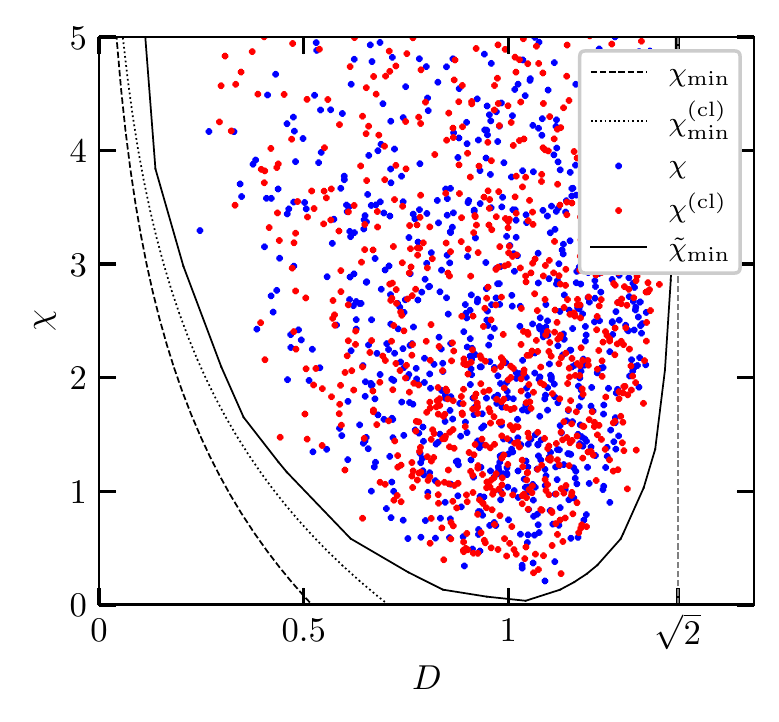}
    \caption{Quantum ($\chi_{\rm min}$) and classical ($\chi_{\rm min}^{\rm (cl)}$) lower bounds on $\chi$, and convex hull ($\tilde{\chi}_{\rm min}$) of $10^6$ random Gaussian states (1000 states plotted as colored circles). Convex hull is the same for quantum and classical sample states.}
\label{fig:random-chi-D}
\end{figure}

A sample of random two-mode Gaussian states is shown in Fig.~\ref{fig:random-chi-D} and compared to Eqs.~\eqref{bipartite-quantum-chi-D-bound} and \eqref{bipartite-classical-chi-D-bound}. Evidently there is a region between the bounds where there may exist states exhibiting a \emph{quantum advantage}, although such states are not present in our random sample. For example, we see that for $D = 0.5$, the classical bound requires $\chi$ to be almost unity, whereas the more permissive quantum bound requires $\chi$ to be only slightly greater than zero. If such states exist with $\chi$ much less than unity for $D = 0.5$, the quantum analysis yields a lower cost to achieve the same degree of synchronization. 

As stated earlier, we assume the system to begin in the uncoupled equilibrium state $\hat{\rho}_0$, and we quantify the departure from this \emph{particular} equilibrium state by introducing a parameter
\begin{equation}
\label{L-q-def}
L= S(\hat{\rho}\| \hat{\rho}_0).
\end{equation}
Unlike $\chi$ that quantifies the distance of the reduced state $\hat{\rho}$ from all possible Gibbs states and \emph{minimizes} over the energy, the parameter $L$ measures the distance with respect to only one specific Gibbs state given by the initial condition. Given the distance measures $\chi$ and $L$ the following chain of inequalities follows,
\begin{equation}
\label{L-chi-lambda-bounds}
L\geq \chi \geq \Lambda.
\end{equation}
Above $\Lambda = \min_{\hat{\sigma}\in \Omega} S(\hat{\rho}\| \hat{\sigma})$ is the relative entropy of entanglement with $\Omega$ being the set of all separable states of the system~\cite{piani_relative_2009}. 
By assumption, $L = \chi = 0$ at time $t=0$. Therefore as a consequence of Eq.~\eqref{N-mode-quantum-chi-D-bound}, for the system to synchronize in time $\tau_s$ to distance $D_s$ at time $\tau_s$ we must have
\begin{equation}
\label{L-sync-requirement}
    L \geq \chi_\text{min}(N, \kappa, D_s).
\end{equation}

Also note that \cite{noauthor_see_nodate} $\dot{L} =\beta \dot{E} -  \dot{S}$, so if we write the first law of thermodynamics as $\dot{E} = \dot{W} - \dot{Q}$, and the second law as $\dot{S} + \beta \dot{Q} \geq 0$~\cite{callen_thermodynamics_1960,Deffner10,Deffner11,landi_irreversible_2021, deffner_quantum_2019}, we have
\begin{equation}
    \dot{W} \geq \frac{1}{\beta}\dot{L}.
\end{equation}
Above $S = -\tr{\hat{\rho}\ln\hat{\rho}}$ is the von Neumann entropy, not to be confused with the relative entropy which will always appear with arguments $S(\cdot \| \cdot)$. Moreover, since $W(0) = L(0)=0$ it follows that $W \geq L$, so we have a lower bound on the amount of work required for synchronization, which is our first main result
\begin{equation}
\label{quantum-work-lower-bound}
W \geq \frac{1}{\beta}\, \chi_\text{min}(N, \kappa, D),
\end{equation}
where $\chi_\text{min}(N,\kappa,D)$ is the right hand side of Eq.~\eqref{N-mode-quantum-chi-D-bound}. Interestingly, the quantity $\chi_\text{min}$ is asymptotically linear in $N$, indicating that the work requirement for synchronization is extensive. The minimum asymptotic work cost
\begin{equation}
    \chi_\text{min}^{\infty} = - \ln \left( \frac{D^2}{8 \kappa}\right) N,
\end{equation}
such that
\begin{equation}
    \lim_{N \to \infty}\frac{\chi_\text{min}}{\chi_\text{min}^{\infty}} = 1.
\end{equation}
Classically, such an asymptotic cost
\begin{equation}
    \chi^{\text{(cl)}\infty}_\text{min} = - \ln \left( \frac{D^2}{2 \kappa}\right) N,
\end{equation}
is lower than the quantum case indicating that in the limit of many oscillators, the thermodynamic costs of synchronizing classical systems will always be lower than synchronizing equivalent quantum systems. However for small values of $N$ the asymptotic expressions are invalid and the classical synchronization cost may be less, and we leave the full investigation of such cases to future work.

\begin{figure}
    \centering
    \includegraphics[width=.48\textwidth]{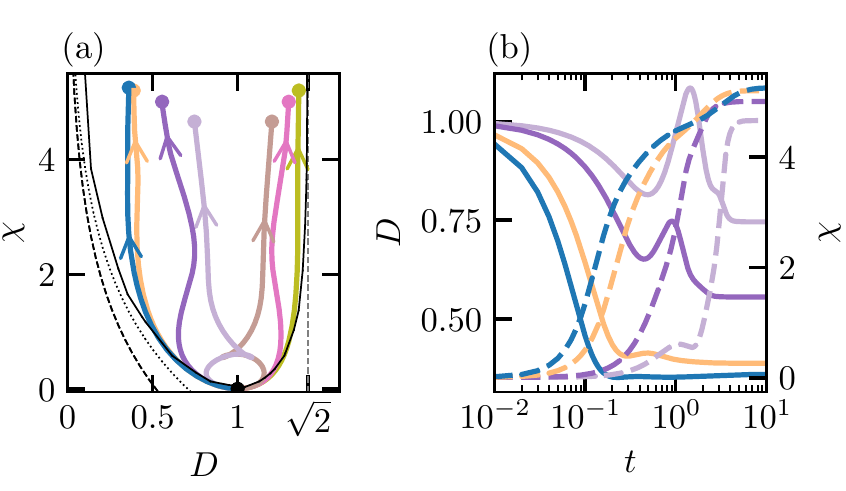}
    \caption{Trajectories in the $\chi$-$D$ plane with $k$ values (from left to right) $k=5$, $k=3$, $k=1$, $k=0.5$, $k=-0.5$, $k=-1$, $k=-3$ are shown in (a) Classical (dotted line) and quantum (dashed line) lower bounds on $\chi$, and convex hull of $10^6$ random Gaussian states (solid line). Time-dependence of $D$ (solid lines) and $\chi$ (dashed lines) with $k$ values (solid lines bottom to top, dashed lines top to bottom) $k=5$, $k=3$, $k=1$, $k=0.5$ in (b) The frequencies of the two oscillators are $\omega_1 = 2\pi$, $\omega_2 = 3\pi$. Both bath temperatures are set to $T=20$. The coupling strengths $\gamma_+= 0.001$ and $\gamma_-$ is determined via $\beta$ and local detailed balance condition as described in the supplemental material~\cite{noauthor_see_nodate}.} 
    \label{fig2}
\end{figure}

\paragraph{Rate of quantum synchronization}

The rate of evolution of $L$ can be found directly from the master equation \eqref{full-QME} \cite{noauthor_see_nodate}, and is given by
\begin{equation}
\label{Ldot_quantum}
\dot{L} =2\, \text{tr}\left\{(\hat{H}_c - \braket{\hat{H}_c}) \hat{\rho}\ln \hat{\rho}\right\} +2\beta_0\mathcal{C}_{CE} - \sigma_0.
\end{equation}
where $\mathcal{C}_{CE} = \frac{1}{2}\braket{\hat{H}_c \hat{H}_0} -\braket{\hat{H}_c}\braket{\hat{H}_0}$ is the covariance of $\hat{H}_0$ and $\hat{H}_c$, and $\sigma_0= -\tr{\mathcal{D}[\hat{\rho}](\ln \hat{\rho} - \ln \hat{\rho}_0)}$ is the irreversible entropy production~\cite{Thingna17} of the uncoupled system, which is always non-negative~\cite{spohn_entropy_1978, breuer_theory_2002}. In the case that $\hat{H}_c$ is a unbounded operator, $\dot{L}$ can be bounded from above,
\begin{equation}
\label{qsl-unbounded}
\begin{split}
   \dot{L} &\leq 2 \Delta_C \sqrt{\mathcal{E}+S_{G;E}^2}\\
   &\quad+2\beta_0\sqrt{ \Delta_E^2 \Delta_C^2 -  \frac{1}{2} \left|\braket{[\hat{H}_0, \hat{H}_c]}\right|^2} - \sigma_0.
\end{split}   
\end{equation}
Here, we use the von Neumann entropy of the Gibbs state, $S_{G;E}=-\text{tr}\{\hat{\rho}_{G;E} \ln \hat{\rho}_{G;E}\}$, as well as the capacity of entanglement~$\mathcal{E} = \tr{\hat{\rho} (\ln \hat{\rho})^2} - S^2 $, which is the second moment of surprisal~\cite{shrimali_capacity_2022}.

In the special case where $\hat{H}_c$ is a bounded operator, we may instead use the alternative bound given in the supplemental material \cite{noauthor_see_nodate} where the capacity of entanglement does not appear. A classical limit of the master equation \eqref{full-QME} can be derived by identifying a quantum phase space distribution with a classical probability density \cite{schleich_quantum_2001, louisell_quantum_1990, graefe_classical_2010}, and a corresponding classical bound (see supplemental material~\cite{noauthor_see_nodate}) reads
\begin{equation}
\label{csl-unbounded}
\begin{split}
    \dot{L}^\text{(cl)} &\leq \beta_0 \braket{\nabla H_c \cdot \nabla H_0} -\braket{\nabla^2 H_c}\\
    &\quad + 2\Delta_C \braket{\ln(f)^2} +2\beta_0\Delta_C \Delta_E - \sigma_0.
\end{split}
\end{equation}

Equation~\eqref{csl-unbounded} differs from Eq.~\eqref{qsl-unbounded} because of the presence of the geometric terms associated with phase space flow \cite{daems_entropy_1999}, as well as the absence of the commutator.  The interpretation of Eq.~\eqref{qsl-unbounded} is as follows: the final term $\sigma_0$ is the rate of irreversible entropy production \cite{spohn_entropy_1978} which one would obtain in the absence of coupling, and this term will always be negative. Therefore the other terms must have a net positive effect larger than this entropy production rate in order for synchronization to occur. The first term is of the form of a quantum speed limit with the additional factor of entropy. This term unsurprisingly implies that stronger non-Hermitian coupling leads to faster synchronization. The second term is reminiscent of the Mandelstam-Tamm quantum speed limit \cite{tamm_uncertainty_1991}, and involves the second moments of both the Hermitian and anti-Hermitian parts of the Hamiltonian, however there is a penalty that scales with the square of their commutator. This term arises from the uncertainty relation \cite{nielsen_quantum_2000}, and can be explained by the fact that synchronization is most effective when there is a large correlation between the observables corresponding to the Hermitian and anti-Hermitian parts of the Hamiltonian.

\begin{figure}
    \centering
    \includegraphics[width=.48\textwidth]{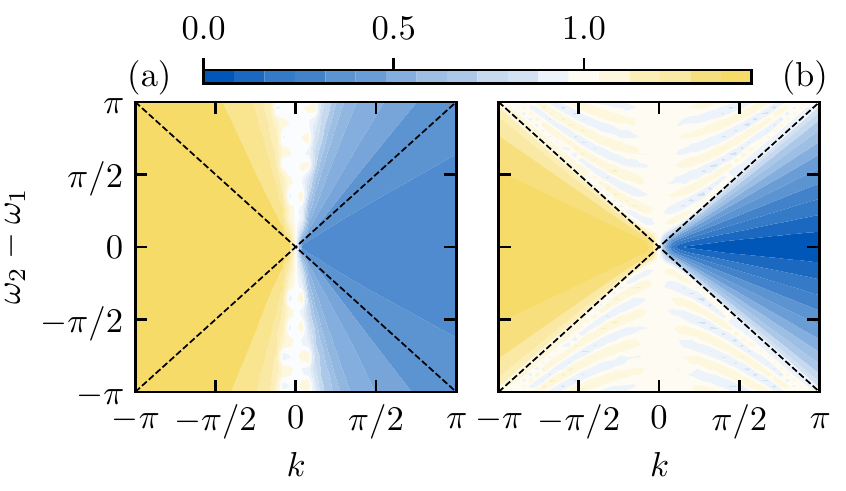}
    \caption{Distance measure $D$ at time $t=10$ as a function of $k$ and $\omega_2-\omega_1$ for quantum (a) and classical (b) evolution. Dashed lines are $k = \pm (\omega_2 - \omega_1)$. Classical synchronous regime is for $k>\left|\omega_2-\omega_1\right|$ (within the dashed lines), whereas for a quantum system the synchronizing regime extends well beyond the classical bounds. All other parameters are same as Fig.~\ref{fig2}.}
    \label{fig3}
\end{figure}

\paragraph{Coupled waveguide model}

Yang et al.~\cite{yang_anti-_2017} use an adiabatic elimination procedure to reduce a system of three coupled waveguides to a system of two waveguides with an effective anti-$\mathcal{P}\mathcal{T}$ symmetric Hamiltonian with the non-Hermitian coupling of the form 
\begin{equation}
\label{dimer-coupling}
    \hat{H}_c = \frac{k}{2}(\hat{a}_1^\dag \hat{a}_2 + \hat{a}_2^\dag \hat{a}_1). 
\end{equation}
In addition to this controlled dissipative interaction, we consider the two modes in contact with heat reservoirs, resulting in a master equation of the form of Eq.~\eqref{full-QME} with four jump operators $\hat{L}_{j+} = (\hat{a}_j^\dag)^2$, $\hat{L}_{j-} = \hat{a}_j^2$. Results of the simulated quantum master equation \eqref{full-QME} with the coupling \eqref{dimer-coupling} are shown in Figs.~\ref{fig2} and \ref{fig3}. Interestingly, in Fig.~\ref{fig2}(a) we see that the trajectories in the $\chi$-$D$ plane are generally confined within the convex hull of randomly generated Gaussian states. However, this is not surprising given that the coupling is bilinear, and the initial state is Gaussian, so the evolved state should be Gaussian as well \cite{serafini_quantum_2023}. In Fig. \ref{fig2}(b) we see the time evolution of $D$ and $\chi$, and in particular it is notable that $\chi$ is almost always monotonically increasing whereas $D$ has more significant reversals in direction. It can also be clearly seen from Fig. \ref{fig2}(a) that $\chi$ is generally increasing even when $D$ does not change appreciably, meaning that work will be wasted in such cases where $k$ is not large enough for synchronization  to occur.

In the classical limit, this dynamics reduces to
\begin{equation}
\label{stuart-landau-1}
    \dot{z}_1 = \left[ \frac{k}{2} - i \omega_1 - \gamma_1 \left| z_1\right|^2\right] + \frac{k}{2} (z_2 - z_1) - i 2 \sqrt{\gamma_1}\xi(t),
\end{equation}
\begin{equation}
\label{stuart-landau-2}
    \dot{z}_2 = \left[ \frac{k}{2} - i \omega_2 - \gamma_2 \left| z_2\right|^2\right] + \frac{k}{2} (z_1 - z_2) - i 2 \sqrt{\gamma_2}\xi(t),
\end{equation}
where we use a complex phase space representation $z_j = (x_j + ip_j)/\sqrt{2}$ and $\xi(t)$ denotes an idealized delta-correlated noise process \cite{noauthor_see_nodate}. Note that Eqs.~\eqref{stuart-landau-1} and \eqref{stuart-landau-2} describe a pair of coupled Stuart-Landau oscillators with amplitude-dependent noise \cite{bi_explosive_2014, selivanov_adaptive_2012}. We find that the classical Stuart-Landau system displays synchronization in the regime $k \geq |\omega_2 - \omega_1|$ (see Fig. \ref{fig3} and supplemental material \cite{noauthor_see_nodate}). The corresponding quantum system is also synchronous for $k \geq |\omega_2-\omega_1|$, and synchronizes for smaller values of $k$ than the classical system \cite{noauthor_see_nodate}. This presents clear evidence of a quantum advantage that leads to a wider regime of synchronization well beyond the classical case. 

\paragraph{Concluding remarks}

In this work is that we have found the minimal amount of work required for synchronization of an arbitrary number of oscillators, as well as bounded the speed at which synchronization may occur. Our numerical results for the model of a anti-$\mathcal{PT}$-symmetric Hamiltonian serve to illustrate an experimentally realizable system where this process may occur. Our analysis is formulated mostly in terms of informational quantities, and therefore allows for an information-theoretic interpretation of the synchronization process as a form of communication. This connection could be made more explicit in future work by relating our measure of synchronization to mutual information. It remains to be seen whether there are states that display a quantum advantage, in the sense of Fig.~\ref{fig:random-chi-D}, and a related goal for future work is to understand the apparent quantum advantage displayed in synchronization of the dimer model (see Fig.~\ref{fig3}). 

\begin{acknowledgments}
J.T. acknowledges support from the Institute for Basic Science in South Korea (IBS-R024-Y2). The authors would like to thank M. Rohith for the useful discussions. S.D. acknowledges support from  the U.S. National Science Foundation under Grant No. DMR-2010127 and the John Templeton Foundation under Grant No. 62422.
\end{acknowledgments}
\bibliography{refs}

\pagebreak
\onecolumngrid
\section*{Supplemental material for\\ ``Time and Energy Requirements for Quantum Synchronization''}

\section{Classical Synchronization Measure}
\label{classical-sync-measure-app}
We set $\hbar=k_B=1$ in what follows. Let $\bold{r} = (x_1, p_1, \dots x_N, p_N)^T$ be the dimensionless phase space coordinates, in terms of which the energy of the uncoupled system is
\begin{equation}
    \label{classical-H0}
    H_0(\bold{r}) = \frac{1}{2}\sum_{j=1}^N \omega_j (x_j^2 + p_j^2).
\end{equation}
The classical Gibbs distribution is \cite{callen_thermodynamics_1960}
\begin{equation}
\label{classical-gibbs-def}
f_{G;E}(\bold{r}) = \frac{1}{(2\pi)^N\sqrt{|\Sigma_{G;E}|}}\exp\left(- \frac{1}{2}\bold{r}^T \Sigma_{G;E}^{-1} \bold{r}\right) = \frac{1}{Z_E} \exp\left(-  \beta _E H_0(\bold{r})\right),
\end{equation}
where $Z_E = (2 \pi)^N \sqrt{\left| \Sigma_{G;E}\right|}$ is the canonical partition function. The average energy is then
\begin{equation}
E = \frac{1}{2}\sum_{j=1}^N \omega_j ( \braket{x_j^2} + \braket{p_j^2}) = \int d^{2N} \bold{r} f(\bold{r}) H_0(\bold{r}).
\end{equation}
We define the parameter $\chi^\text{(cl)}$ to be the relative entropy from the distribution $f$ to the equilibrium manifold of the uncoupled system,
\begin{equation}
    \chi^\text{(cl)} = \min_{U} S^\text{(cl)}(f\| f_{G;U}), 
\end{equation}
where we use the classical relative entropy \cite{kullback_information_1951}
\begin{equation}
\label{classical-rel-ent-def}
    S^\text{(cl)}(f\| g) = \int d^{2N} \bold{r} f(\bold{r}) (\ln f(\bold{r}) - \ln g(\bold{r})).
\end{equation}
We first show that
\begin{equation}
    \label{classical-chi-minimization}
    \chi^\text{(cl)} = S^\text{(cl)}(f\| f_{G;E}).
\end{equation}
The relative entropy to the Gibbs distribution is given by
\begin{align}
     S^\text{(cl)}(f\| f_{G;U})& =  \int d^{2N} \bold{r} f(\bold{r}) (\ln f(\bold{r}) - [ - \beta_{U}H_0(\bold{r})- \ln Z_{U}])\\
    & = - S^\text{(cl)}+ \beta_{U} E  + \ln Z_{U} \\
    &=- S^\text{(cl)}+ \beta_{U} E  + S^\text{(cl)}_{G;U} - \beta_U U\\
    & =- S^\text{(cl)}+ \beta_{U} (E - U)  + S^\text{(cl)}_{G;U} \\,
\end{align}
where we used the thermodynamic identity $\ln Z_U = S_{G;U}^\text{(cl)} - \beta_U U$ \cite{callen_thermodynamics_1960}. Only the last two terms depend on $U$, so we minimize these terms with  respect to $U$. We assume that temperature is monotonically increasing with energy, which is true for our harmonic oscillator Hamiltonian. Note because $\beta_U = \partial_U S_{G;U}^\text{(cl)}$, this implies that for all $U$, $E$
\begin{equation}
    S^\text{(cl)}_{G;E} \leq \beta_U(E- U) + S^\text{(cl)}_{G;U},
\end{equation}
so for fixed $E$ and arbitrary $U$,
\begin{equation}
    S^\text{(cl)}(f\|f_{G;E}) \leq S^\text{(cl)}(f\|f_{G;U}).
\end{equation}
By the above reasoning, when temperature is monotonically increasing with energy, the relative entropy to the equilibrium manifold is minimized when $U=E$, completing the proof of Eq. \eqref{classical-chi-minimization}. This argument is not valid in the case of negative heat capacity; although such systems exist they are rare \cite{schmidt_negative_2001} and our coupled harmonic oscillator system is not among them. The quantity $\chi^\text{(cl)}$ can then be expressed as
\begin{equation}
    \chi^\text{(cl)} = S^\text{(cl)}_{G;E} - S^\text{(cl)},
\end{equation}
and so $\chi^\text{(cl)}$ can be lower bounded by obtaining an upper bound for $S^{\text{(cl)}}$. This can be done using the fact that for a given covariance matrix $\Sigma$, a multivariate normal distribution maximizes the Shannon entropy \cite{cover_elements_2006}, which the maximal value being
\begin{equation}
    S^\text{(cl)}_{\Sigma} = \frac{1}{2}\ln \left(\left| 2 \pi e \Sigma \right|\right).
\end{equation}
For the system of coupled harmonic oscillators, the Gibbs distribution $f_{G;E}$ is itself is a multivariate normal distribution having covariance matrix $\Sigma_{G;E}$. Therefore
\begin{equation}
    \chi \leq S^\text{(cl)}_{G;E} - S^\text{(cl)}_{\Sigma} =  \ln \left(\frac{\sqrt{\left|\Sigma_{G;E} \right|}}{\sqrt{\left| \Sigma\right|}}\right)
\end{equation}
The covariance matrix of the Gibbs distribution is diagonal, and the quadrature variances are
\begin{equation}
    \braket{x_j^2}=\braket{p_j^2} = \frac{E}{N \omega_j}.
\end{equation}
Let $\omega_m$ and $\omega_M$ be respectively the smallest and largest of the frequencies $\omega_j$, and $\kappa = \omega_m/\omega_M$ be their ratio, in terms of which we can obtain the following bound
\begin{align}\sqrt{\left| \Sigma_{G;E}\right|}& = \prod_{j=1}^N \frac{E}{N\omega_j } \geq \left(\frac{E}{N\omega_M } \right)^{N} \\
& \geq 
\left(\frac{\omega_m \braket{\bold{r}^2}}{2N\omega_M } \right)^{N} = \left(\frac{\kappa\braket{\bold{r}^2}}{2N} \right)^{N} \\
& = \left(\frac{\kappa \braket{\overline{\bold{r}^2}}}{2} \right)^{N},
\end{align}
where on the last line we have used the quantity $\braket{\overline{\bold{r}^2}} = \braket{\bold{r}^2}/N$.
We make a change of basis to coordinates $c_1 \dots c_{2N}$, which represent an orthonormal basis. In particular, we choose
\begin{equation}
    c_1 = \frac{1}{\sqrt{N}} \sum_j x_j = \sqrt{N} \bar{x}, \: \: \:
    c_2 = \frac{1}{\sqrt{N}} \sum_j p_j = \sqrt{N} \bar{p},
\end{equation}
where $\bar{x} = N^{-1}\sum_j x_j$ and $\bar{p} = N^{-1}\sum_j p_j$. The remaining coordinates $c_3\dots c_{2N}$ may then be chosen using the Gram-Schmidt process, for example \cite{lay_linear_2012}. The determinant may be evaluated in the new basis. Using Fischer's inequality \cite{thompson_determinantal_1961} and the inequality of arithmetic and geometric means (AM/GM) \cite{steele_cauchy-schwarz_2004}, we find
\begin{align}
\sqrt{\left|\Sigma\right|}  &\leq \prod_{j=1}^{2N}\braket{c_j^2} =\sqrt{\braket{c_1^2} \braket{c_2^2}}  \left[\left(\prod_{j=3}^{2N} \braket{c_j^2}\right)^{1/(2N-2)}\right]^{N-1} \\
&\leq \frac{1}{2}(\braket{c_1^2}+\braket{c_2^2}) \left(\frac{1}{2N - 2}\sum_{j=3}^{2N} \braket{c_j^2}\right)^{N-1} \\
&\leq 
 \frac{1}{2(2 N - 2)^{N-1}}\braket{\bold{r}^2} \left(\braket{\bold{r}^2} - c_1^2 - c_2^2\right)^{N-1} \\
 &= 
 \frac{N}{2(2 N - 2)^{N-1}}\braket{\overline{\bold{r}^2}} \left(N\braket{\overline{\bold{r}^2}} - N \braket{\bar{\bold{r}}^2}\right)^{N-1} \\
 & = \frac{N^N(N-1)}{(2N-2)^N}
 \braket{\overline{\bold{r}^2}} \left(\braket{\overline{\bold{r}^2}} -  \braket{\bar{\bold{r}}^2}\right)^{N-1}
\end{align}
So 
\begin{align}
    \chi^\text{(cl)} &\geq \ln \left( \frac{\sqrt{\left|\Sigma_{G;E}\right|}}{\sqrt{\left|\Sigma\right|}}\right) \\
    & \geq \ln \left(\frac{(2N -2)^N}{N^N(N-1)}\frac{ \left(\kappa \braket{\overline{\bold{r}^2}}/2\right)^{N}}{
 \braket{\overline{\bold{r}^2}} \left(\braket{\overline{\bold{r}^2}} -  \braket{\bar{\bold{r}}^2}\right)^{N-1}} \right) \\
 & = -\ln \left(\frac{N^N}{(N-1)^{N-1}\kappa^N}\frac{\left(\braket{\overline{\bold{r}^2}} -  \braket{\bar{\bold{r}}^2}\right)^{N-1}}{\braket{\overline{\bold{r}^2}}^{N-1}} \right) \\
 &= 
 -\ln \left(\frac{N^N}{(N-1)^{N-1}\kappa^N}\left(\frac{1}{2} D^2 \right)^{N-1}\right) \\
 &= 
 -2(N-1)\ln \left(\sqrt{\frac{(N/\kappa)^{N/(N-1)}}{2(N-1)}}D\right) 
\end{align}
Note that for $N=2$ this reduces to
\begin{equation}
    \chi^{\text{(cl)}} \geq -2\ln \left( \frac{\sqrt{2} D}{\kappa}\right).
\end{equation}
In the limit of large $N$, asymptotically the lower bound is given by
\begin{equation}
\chi^\text{(cl)}_\text{min:as} = - 2\ln \left( \frac{ D}{ \sqrt{2\kappa}}\right) N,
\end{equation}
To be precise,
\begin{equation}
    \lim_{N \to \infty}\frac{\chi_\text{min}^\text{(cl)}}{\chi_\text{min:as}^\text{(cl)}} = 1.
\end{equation}

\section{Quantum synchronization measure}
\label{bipartite-quantum-sync-measure-appendix}
Let $\hat{\bold{r}} = (\hat{x}_1, \hat{p}_1, \dots \hat{x}_N, \hat{p}_N)^T$ be the dimensionless quadrature operators satisfying the canonical commutation relations
\begin{equation}
\label{canonical-commutation-relations}
[\hat{x}_j,\hat{p}_k] = i\delta_{jk},
\end{equation}
and $\hat{a}_j, \hat{a}_j^\dag$ are the corresponding annihilation and creation operators. The Hamiltonian of the uncoupled system is
\begin{equation}
    \label{quantum-H0}
    \hat{H}_0 =\frac{1}{2}\sum_{j=1}^N \omega_j (\hat{x}_j^2 + \hat{p}_j^2) = \sum_{j=1}^N\omega_j\left(\hat{n}_j + \frac{1}{2}\right).
\end{equation}
The Gibbs state is \cite{deffner_quantum_2019}
\begin{equation}
    \label{quantum-gibbs-def}
    \hat{\rho}_{G;U} = \frac{\exp(-\beta_E \hat{H}_0)}{\text{tr}\{\exp(-\beta_E \hat{H}_0)\}} = \frac{1}{Z_U}\exp(-\beta_E \hat{H}_0) ,
\end{equation}
where $Z_E = \text{tr}\{\exp(-\beta_E \hat{H}_0)\}$ is the canonical partition function. In the quantum case, $\chi$ is expressed in terms of the quantum relative entropy
\begin{equation}
\label{quantum-chi-minimization}
    \chi = \min_U S(\hat{\rho}\|\hat{\rho}_{G;U}),
\end{equation}
where $S(\hat{\rho}\|\hat{\sigma}) = \text{tr}\{\hat{\rho} (\ln \hat{\rho} - \ln \hat{\sigma})\}$ \cite{vedral_role_2002}. We proceed in a similar way to the classical case, and write the relative entropy to the Gibbs state as
\begin{align}
     S(\hat{\rho}\| \hat{\rho}_{G;U})& =  \text{tr}\{\hat{\rho} (\ln \hat{\rho} - \ln \hat{\rho}_{G;U})\}\\
    & = - S+ \beta_{U} E  + \ln Z_{U} \\
    &=- S+ \beta_{U}E  + S_{G;U} -  \beta_{U} U  \\
    & =- S + \beta_{U}(E - U) +S_{G;U}.
\end{align}
Again, only the last two terms depend on $U$, and we minimize these terms. As in the classical case, temperature is monotonically increasing with energy for the quantum harmonic oscillator. By the same reasoning as in the classical case, we have for all $U$, $E$
\begin{equation}
    S_{G;E} \leq \beta_U(E- U) + S_{G;U},
\end{equation}
so for fixed $E$ and arbitrary $U$,
\begin{equation}
    S(f\|f_{G;E}) \leq S(f\|f_{G;U}),
\end{equation}
which completes the proof of Eq. \eqref{quantum-chi-minimization}.  It follows that $\chi$ can be expressed as
\begin{equation}
    \chi = S_{G;E} - S,
\end{equation}
and we will proceed by finding a lower bound for $S_{G;E}$ and an upper bound for $S$. The entropy of a single-mode Gaussian state is \cite{serafini_symplectic_2004}
\begin{equation}
    S_1 = \frac{1-\mu}{2\mu}\ln \left( \frac{1+\mu}{1-\mu}\right) - \ln \left( \frac{2\mu}{1+\mu}\right),
\end{equation}
where 
\begin{equation}
\mu =\frac{1}{2\sqrt{\left|\Sigma\right|}}.
\end{equation}
Useful upper and lower bounds on this quantity are
\begin{equation}
  \ln(2\sqrt{\left| \Sigma \right|})\leq   S_1 \leq \ln( e \sqrt{\left| \Sigma \right|}). 
\end{equation}
The Gibbs state of the uncoupled Hamiltonian is a tensor product state, so its von Neumann entropy is additive, meaning
\begin{equation}
    S_{G;E} \geq \sum_j \ln \left(2 \sqrt{\left|\Sigma_j \right|}\right) = \ln \left(2^N\sqrt{\left|\Sigma_{G;E} \right|}\right),
\end{equation}
where we have also used the fact that the covariance matrix is diagonal in the Gibbs state \cite{serafini_quantum_2023}. By the same steps as we used in the classical case, we have
\begin{equation}
\sqrt{\left| \Sigma_{G;E} \right|} \geq \left(\frac{\kappa \braket{\overline{\bold{r}^2}}}{2} \right)^{N},
\end{equation}
so
\begin{equation}
    S_{G;E} \geq N\ln \left(\kappa \braket{\overline{\bold{r}^2}} \right).
\end{equation}
We also require an upper bound on $S$. This can be obtained by noting that the von Neumann entropy is always sub-additive, so
\begin{equation}
    S \leq \sum_j \ln\left(e \sqrt{\left| \Sigma_j\right|}\right) = \ln \left(e^N \prod_{j=1}^N \sqrt{\left|\Sigma_j\right|}\right).
\end{equation}
Again we may make a change of basis to the operators $\hat{c}_1 \dots \hat{c}_{2N}$, which are defined in the same way as in the classical case, but have been promoted to operators. By Fischer's inequality,
\begin{equation}
    S  \leq \ln \left(e^N \prod_{j=1}^{2N} \braket{c_j^2}\right),
\end{equation}
and so the same steps in the classical case give the result
\begin{equation}
   S \leq \ln\left( e^N\frac{N^N(N-1)}{(2N-2)^N}
 \braket{\overline{\bold{r}^2}} \left(\braket{\overline{\bold{r}^2}} -  \braket{\bar{\bold{r}}^2}\right)^{N-1}
\right)
\end{equation}
We then have the lower bound for $\chi$,
\begin{equation}
    \chi \geq \ln \left(\frac{(2N -2)^N}{e^N N^N(N-1)}\frac{ \left(\kappa \braket{\overline{\bold{r}^2}}\right)^{N}}{
 \braket{\overline{\bold{r}^2}} \left(\braket{\overline{\bold{r}^2}} -  \braket{\bar{\bold{r}}^2}\right)^{N-1}} \right) \\,
\end{equation}
and we see that the argument of the log differs from the classical case by a factor of $(2 /e)^N$. Therefore we have the lower bound
\begin{equation}
\chi_\text{min} = \chi_\text{min}^\text{(cl)} - N (1 - \ln 2 ).
\end{equation}
Explicitly,
\begin{equation}
\label{chi-min-quantum}
 \chi \geq -2(N-1)\ln \left(\sqrt{\frac{(e N/2\kappa)^{N/(N-1)}}{2(N-1)}}D\right) .
\end{equation}
For $N=2$ this reduces to
\begin{equation}
    \chi \geq -2\ln \left( \frac{e D}{\sqrt{2}\kappa}\right).
\end{equation}
In the limit of large $N$, the lower bound in Eq. \eqref{chi-min-quantum} is asymptotically given by
\begin{equation}
\chi_\text{min:as} = - 2\ln \left( \frac{\sqrt{e} D}{2 \sqrt{\kappa}}\right) N,
\end{equation}
And again, we have
\begin{equation}
    \lim_{N \to \infty}\frac{\chi_\text{min}}{\chi_\text{min:as}} = 1.
\end{equation}

\section{Quantum synchronization speed limit}
\label{qsl-supp}
We assume that the generator of dynamics has the form
\begin{equation}
    \label{full-QMES}
\mathcal{L}[\hat{\rho}] = -i[\hat{H}_0, \hat{\rho}] + \{\hat{H}_c, \hat{\rho}\} - 2\braket{\hat{H}_c}\hat{\rho} + \mathcal{D}[\hat{\rho}],
\end{equation}
where
\begin{equation}
    \label{dissipator-def}
    \mathcal{D}[\hat{\rho}]=\sum_i \hat{F}_i \hat{\rho} \hat{F}_i^\dag - \frac{1}{2}\left\{ \hat{F}_i^\dag \hat{F}_i,\hat{\rho}\right\}.
\end{equation}
We also define
\begin{equation}
    \mathcal{L}_0[\hat{\rho}] = -i[\hat{H}_0,\hat{\rho}]+\mathcal{D}[\hat{\rho}],
\end{equation}
\begin{equation}
        \mathcal{L}_c[\hat{\rho}] = \{\hat{H}_c,\hat{\rho}\}- \braket{\hat{H}_c}\hat{\rho}.
\end{equation}
Here we derive a bound on the rate at which the quantity $L$ can change, which can then be used to determine the minimal time for quantum synchronization. Recall that $L$ is defined as
\begin{equation}
    \label{L-quantum-2}
    L = \text{tr}\{\hat{\rho}(\ln \hat{\rho} - \ln \hat{\rho}_0)\},
\end{equation}
and its time-derivative can be expressed as \cite{breuer_theory_2002}
\begin{equation}
    \dot{L} = \text{tr}\{\mathcal{L}[\hat{\rho}](\ln \hat{\rho} - \ln \hat{\rho}_0)\}.
\end{equation}
Using the definition $\mathcal{L}= \mathcal{L}_0 + \mathcal{L}_c$, we have
\begin{equation}
    \dot{L} = \dot{L}_0 + \dot{L}_c,
\end{equation}
where
\begin{equation}
    \dot{L}_0 = \text{tr}\{\mathcal{L}_0[\hat{\rho}](\ln \hat{\rho} - \ln \hat{\rho}_0)\},
\end{equation}
and
\begin{equation}
    \dot{L}_c = \text{tr}\{\mathcal{L}_c[\hat{\rho}](\ln \hat{\rho} - \ln \hat{\rho}_0)\}.
\end{equation}
Evaluating $\dot{L}_0$, we find
\begin{align}
    \dot{L}_0 & = -i\text{tr}\{[\hat{H}_0,\hat{\rho}](\ln \hat{\rho} - \ln \hat{\rho}_0)\} + \text{tr}\{\mathcal{D}[\hat{\rho}](\ln \hat{\rho} - \ln \hat{\rho}_0)\}\\
    &= 
-i\text{tr}\{\hat{H}_0[\hat{\rho},\ln\hat{\rho}] - [\ln \hat{\rho}_0,\hat{H}_0]\hat{\rho}_0]\} + \text{tr}\{\mathcal{D}[\hat{\rho}](\ln \hat{\rho} - \ln \hat{\rho}_0)\}\\
&=
\text{tr}\{\mathcal{D}[\hat{\rho}](\ln \hat{\rho} - \ln \hat{\rho}_0)\},
\end{align}
where we used the fact that $[\hat{\rho},\ln \hat{\rho}]=0$ and $[\hat{H}_0, \ln \hat{\rho}_0] = 0$. Due to a result by Spohn \cite{spohn_entropy_1978}, the above can be equated to the negative irreversible entropy production rate of the uncoupled system, which we call $-\sigma_0$
\begin{equation}
    \dot{L}_0 = -\sigma_0.
\end{equation}
We next evaluate $\dot{L}_c$,
\begin{align}
    \dot{L}_c & = \text{tr}\left\{\left(\{\hat{H}_c,\hat{\rho}\} - 2 \braket{\hat{H}_c} \hat{\rho}\right)(\ln \hat{\rho} - \ln \hat{\rho}_0)\right\}\\
 &= \text{tr}\left\{\left(\{\hat{H}_c,\hat{\rho}\} - 2 \braket{\hat{H}_c} \hat{\rho}\right)\ln \hat{\rho}\right\} - \text{tr}\left\{\left(\{\hat{H}_c,\hat{\rho}\} - 2 \braket{\hat{H}_c}\hat{\rho}\right)\ln \hat{\rho}_0\right\}
 \\
 &= 2\, \text{tr}\left\{(\hat{H}_c - \braket{\hat{H}_c}) \hat{\rho}\ln \hat{\rho}\right\} - \text{tr}\left\{\left(\{\hat{H}_c,\hat{\rho}\} - 2\braket{\hat{H}_c} \hat{\rho}\right)\ln \hat{\rho}_0\right\}
 \\
  &=2\, \text{tr}\left\{(\hat{H}_c - \braket{\hat{H}_c}) \hat{\rho}\ln \hat{\rho}\right\} - \text{tr}\left\{\left(\{\hat{H}_c,\hat{\rho}\} - 2\braket{\hat{H}_c} \hat{\rho}\right)(-\beta_0 \hat{H}_0 - \ln Z_0)\right\}
 \\
   &= 2\, \text{tr}\left\{(\hat{H}_c - \braket{\hat{H}_c}) \hat{\rho}\ln \hat{\rho}\right\} + \ln Z_0 \:  \text{tr}\left\{\{\hat{H}_c,\hat{\rho}\} - 2\braket{\hat{H}_c} \hat{\rho}\right\}+\beta_0\:\text{tr}\left\{\left(\{\hat{H}_c,\hat{\rho}\} - 2\braket{\hat{H}_c}  \hat{\rho}\right)\hat{H}_0 \right\}  
 \\
    &= 2\, \text{tr}\left\{(\hat{H}_c - \braket{\hat{H}_c}) \hat{\rho}\ln \hat{\rho}\right\} +\beta_0\:\text{tr}\left\{\left(\{\hat{H}_c,\hat{\rho}\} - 2\braket{\hat{H}_c} \hat{\rho}\right)\hat{H}_0 \right\}  
 \\
     &= 2\, \text{tr}\left\{(\hat{H}_c - \braket{\hat{H}_c}) \hat{\rho}\ln \hat{\rho}\right\}+\beta_0\:\text{tr}\left\{\{\hat{H}_c,\hat{\rho}\} \hat{H}_0 \right\} - 2\beta_0 \braket{\hat{H}_c}\braket{\hat{H}_0}  
 \\
      &= 2\, \text{tr}\left\{(\hat{H}_c - \braket{\hat{H}_c}) \hat{\rho}\ln \hat{\rho}\right\} +\beta_0\:\text{tr}\left\{\{\hat{H}_0,\hat{H}_c\} \hat{\rho} \right\} - 2\beta_0 \braket{\hat{H}_c}\braket{\hat{H}_0}  
 \\
        &= 2\, \text{tr}\left\{(\hat{H}_c - \braket{\hat{H}_c}) \hat{\rho}\ln \hat{\rho}\right\}  +2\beta_0\mathcal{C}_{CE},
 \\
\end{align}
where the covariance $\mathcal{C}_{CE}$ is defined as \cite{serafini_quantum_2023}
\begin{equation}
    \mathcal{C}_{CE} = \frac{1}{2} \braket{\{\hat{H}_0, \hat{H}_c\}} - \braket{\hat{H}_0}\braket{\hat{H}_c}.
\end{equation}
Putting together the expressions for $\dot{L}_0 $ and $\dot{L}_c $, we get
\begin{equation}
\label{L-speed-lim-quantum-appendix}
   \dot{L} =2\, \text{tr}\left\{(\hat{H}_c - \braket{\hat{H}_c}) \hat{\rho}\ln \hat{\rho}\right\} +2\beta_0\mathcal{C}_{CE} - \sigma_0.
\end{equation}
It follows from the above analysis that this can be written as
\begin{equation}
    \dot{L} = \beta_0 \dot{E}_c -  \dot{S}_c - \sigma_0,
\end{equation}
with
\begin{equation}
    \dot{E}_c = \text{tr}\{\mathcal{L}_c[\hat{\rho}] \hat{H}_0\} = 2\mathcal{C}_{CE},
\end{equation}
and
\begin{equation}
    \dot{S}_c = -\text{tr}\{\mathcal{L}_c[\hat{\rho}]\ln \hat{\rho}\} =- 2\text{tr}\left\{(\hat{H}_c - \braket{\hat{H}_c}) \hat{\rho}\ln \hat{\rho}\right\}.
\end{equation}
In order to bound the first term in Eq. \eqref{L-speed-lim-quantum-appendix}, we expand the density operator in its eigenbasis,
\begin{equation}
    \hat{\rho}= \sum_j p_j \ket{\phi_j}\bra{\phi_j},
\end{equation}
so this term takes the form
\begin{equation}
    -\dot{S}_c = 2\sum_j (\braket{\phi_j | \hat{H}_c |\phi_j}-\braket{\hat{H}_c})p_j \ln p_j.
\end{equation}
In the case that $\hat{H}_c$ is a bounded operator, we have in terms of the operator norm
\begin{equation}
   \left| \braket{\phi_j|\hat{H}_c|\phi_j} - \braket{\hat{H}_c} \right|\leq 2\|\hat{H}_c\|,
\end{equation}
leading to the bound
\begin{equation}
    \left|\dot{S}_c\right|  \leq 4 \|\hat{H}_c\|S.
\end{equation}
Finally, we can bound the entropy by the Gibbs entropy of the uncoupled equilibrium state with the same value of $\braket{H_0}$, to give
\begin{equation}
    \left|\dot{S}_c\right|  \leq 4 \|\hat{H}_c\|S_{G;E}.
\end{equation}
If $\hat{H}_c$ is not bounded, as is often the case in the continuous variable setting, we may bound $\dot{S}_c$ using the Cauchy-Schwartz inequality as follows
\bigskip
\begin{align}
   \left| \dot{S}_c\right| &= 2\left|\tr{(\hat{H}_c - \braket{\hat{H}_c})\hat{\rho} \ln \hat{\rho}}\right|\\
    &= 2\left|\tr{(\hat{H}_c - \braket{\hat{H}_c})\sqrt{\hat{\rho}} \sqrt{\hat{\rho}} \ln \hat{\rho}}\right|\\
    &\leq 2\sqrt{ \tr{(\hat{H}_c - \braket{\hat{H}_c})^2\hat{\rho} }} \sqrt{ \tr{\hat{\rho} (\ln \hat{\rho})^2 }}\\
    & =2 \Delta_C \sqrt{\mathcal{E}+S^2}, 
\end{align}
where $\mathcal{E}$ is known as the capacity for entanglement, and is defined as \cite{shrimali_capacity_2022}
\begin{equation}
    \label{entanglement-capacity-appendix}
    \mathcal{E} = \tr{\hat{\rho} (\ln \hat{\rho})^2} - S^2.
\end{equation}
We again use the Gibbs entropy of the uncoupled equilibrium state to upper bound the entropy, giving
\begin{equation}
     \left| \dot{S}_c\right| \leq 2\Delta_C \sqrt{\mathcal{E}+S_{G;E}^2}.
\end{equation}
Using the Schrodinger uncertainty relation \cite{nielsen_quantum_2000}
\begin{equation}
    \Delta_E^2 \Delta_C^2 \geq |\mathcal{C}_{CE}|^2 + \frac{1}{2} \left|\braket{[\hat{H}_0, \hat{H}_c]}\right|^2,
\end{equation}
so
\begin{equation}
    |\mathcal{C}_{CE}| \leq  \sqrt{ \Delta_E^2 \Delta_C^2 -  \frac{1}{2} \left|\braket{[\hat{H}_0, \hat{H}_c]}\right|^2}.
\end{equation}
Using the bounds obtained above, in the case that $\hat{H}_c$ is bounded we find
\begin{equation}
\label{QSL-bounded}
   \dot{L} \leq 4 \|\hat{H}_c\|S_{G;E}+2\beta_0\sqrt{ \Delta_E^2 \Delta_C^2 -  \frac{1}{2} \left|\braket{[\hat{H}_0, \hat{H}_c]}\right|^2} - \sigma_0.
\end{equation}
In the case where $\hat{H}_c$ is unbounded, instead we have
\begin{equation}
\label{QSL-unbounded}
   \dot{L} \leq 2 \Delta_C \sqrt{\mathcal{E}+S_{G;E}^2}+2\beta_0\sqrt{ \Delta_E^2 \Delta_C^2 -  \frac{1}{2} \left|\braket{[\hat{H}_0, \hat{H}_c]}\right|^2} - \sigma_0.
\end{equation}

\section{Classical Master Equation}
\label{CME-appendix}
Graefe et al \cite{graefe_classical_2010} give a recipe for obtaining a classical limit of a non-Hermitian von Neumann equation (without a Lindblad form dissipator). The equation thus obtained is deterministic, and describes the evolution of a system whose initial distribution is a delta function in the phase space, and which remains a delta function over time. In order to find the evolution of a distribution, it is not permissible to simply use the Liouville equation with the drift velocity obtained by Graefe et al, as the total quantum master equation is nonlinear, and therefore we cannot take a linear superposition of the delta function solutions. Here we generalize the arguments from \cite{graefe_classical_2010} to the case of classical probability distributions, and also allow for a Fokker-Planck dissipative term \cite{zwanzig_nonequilibrium_2001}, resulting in a classical master equation analogous to Eq. \eqref{full-QMES}.

We first write the Hamiltonian part of the quantum master equation as
\begin{equation}
\label{quantum-hamiltonian-part}
    \mathcal{L}_H[\hat{\rho}] = -i[\hat{H}_0,\hat{\rho}] +\{\hat{H}_c, \hat{\rho} \}- 2\braket{\hat{H}_c} \hat{\rho}  =     -i[\hat{H}_0,\hat{\rho}] + [\hat{H}_c, \rho] + 2\hat{\rho} \hat{H}_c- 2 \braket{\hat{H}_c} \hat{\rho} .
\end{equation}
We consider the evolution of the quantities $\braket{\hat{a}_j}$ due to the Hamiltonian part only,
\begin{equation}
\label{quantum-hamiltonian-part-avg}
\tr{\mathcal{L}_H[\hat{\rho}]\hat{a}_j} = - i \braket{[\hat{a}_j,\hat{H}_0]} + \braket{[\hat{a}_j, \hat{H}_c]} + 2\braket{ \hat{H}_c \hat{a}_j} -2\braket{\hat{a}_j} \braket{\hat{H}_c}.
\end{equation}
Whereas Graefe et al assume the system is in a coherent state $\ket{\alpha}\bra{\alpha}$, we allow for a general state which can be expressed as a sum of coherent states in the P representation \cite{schleich_quantum_2001},
\begin{equation}
    \hat{\rho} = \int d^{2N} \pmb{\alpha} P(\pmb{\alpha})\ket{\pmb{\alpha}}\bra{\pmb{\alpha}},
\end{equation}
where $P(\pmb{\alpha})$ is known as the P function of the state $\hat{\rho}$. So we have
\begin{align}
    \tr{\mathcal{L}_H[\hat{\rho}]\hat{a}_j} &=
    \int d^{2N} \pmb{\alpha} \,P(\pmb{\alpha})  \left(- i \braket{\pmb{\alpha}|[\hat{a}_j,\hat{H}_0]|\pmb{\alpha}} +  \braket{\pmb{\alpha}|[\hat{a}_j, \hat{H}_c]|\pmb{\alpha}} + 2\braket{\pmb{\alpha}|\hat{H}_c \hat{a}_j|\pmb{\alpha}} - 2\braket{\pmb{\alpha}| \hat{a}_j|\pmb{\alpha}}\braket{\hat{H}_c}\right) \\
    &= 
    \int d^{2N} \pmb{\alpha}\, P(\pmb{\alpha})  \left(- i \braket{\pmb{\alpha}|[\hat{a}_j,\hat{H}_0]|\pmb{\alpha}} +  \braket{\pmb{\alpha}|[\hat{a}_j, \hat{H}_c]|\pmb{\alpha}} + 2\alpha_j \braket{\pmb{\alpha}|\hat{H}_c |\pmb{\alpha}} - 2\alpha_j\braket{\hat{H}_c}\right).
\end{align}
Making use of the formal representation \cite{louisell_quantum_1990, graefe_classical_2010}
\begin{equation}
    [\hat{a}_j, \hat{X}] = \frac{\partial \hat{X}}{\partial \hat{a}_j^\dag}, \: \: \: \:     [\hat{a}_j^\dag, \hat{X}] =- \frac{\partial \hat{X}}{\partial \hat{a}_j} 
\end{equation}
we have
\begin{align}
    \tr{\mathcal{L}_H[\hat{\rho}]\hat{a}_j} &=
    \int d^{2N} \pmb{\alpha}\,  P(\pmb{\alpha})  \left(- i \Braket{\pmb{\alpha}|\frac{\partial \hat{H}_0}{\partial \hat{a}^\dag_j}|\pmb{\alpha}} + \Braket{\pmb{\alpha}|\frac{\partial \hat{H}_c}{\partial \hat{a}^\dag_j}|\pmb{\alpha}} + 2\alpha_j \braket{\pmb{\alpha}|\hat{H}_c |\pmb{\alpha}} - 2\alpha_j\braket{\hat{H}_c}\right)
\end{align}
We now introduce the classical phase space coordinates $\bold{r} = (q_1, p_1, q_2, p_2)^T$, as well as complex coordinates $z_j = (q_j + i p_j)/\sqrt{2}$. The Hamiltonian can be viewed as a function of the quadrature variables $H(\bold{r})$ or of the complex coordinates $H(\bold{z}, \bold{z}^*)$. Because the quantum Hamiltonian is generally represented in terms of the operators $\hat{a}_j$ and $\hat{a}_j^\dag$, we will treat it here as a function of $z_j$ and $z^*_j$, which respectively take the place of $\hat{a}_j$ and $\hat{a}_j^\dag$. To obtain a classical limit, we will replace averages as $\braket{\hat{a}_j} \to \braket{z_j}$, $\braket{\hat{a}_j^\dag}\to \braket{z_j^*}$. The classical probability density function $f$ will be treated as the limiting form of the P function. Finally, an operator which can be expressed as an analytic function $\hat{X}(\hat{a}, \hat{a}^\dag)$ can be cast as a function of complex variable $z$ as $X(z, z)= \braket{z|\hat{X}|z}$, where $\ket{z}$ is the coherent state. With these substitutions, we find
\begin{align}
\label{classical-average-hamiltonian-evo}
   \partial_t \braket{z_j^H} &=
    \int d^{2N} \pmb{z} \, f(\pmb{z})  \left(- i \frac{\partial H_0}{\partial z_j^*}+ \frac{\partial H_c}{\partial z_j^*} +2z_j H_c - 2z_j\braket{H_c}\right).
\end{align}
Note that in the above, the average $\braket{H_c}$ is interpreted as
\begin{equation}
    \braket{H_c} = \int d^{2N} \pmb{z} f(\pmb{z}) H_c (\pmb{z}),
\end{equation}
and so the last term makes the equation nonlinear in $f$. Taking a delta function $f(\pmb{z}) = \delta(\pmb{z} - \pmb{Z})$, we see that the last two terms in Eq. \eqref{classical-average-hamiltonian-evo} cancel, and that we are left with
\begin{equation}
\label{classical-deterministic-hamiltonian-evo}
    \dot{Z_j} = - i \frac{\partial H_0}{\partial z_j^*}+ \frac{\partial H_c}{\partial z_j^*}.
\end{equation}
 Looking at the first term, we see that
\begin{align}
    -i \frac{\partial H_0}{\partial z_j^*} &= -i \left(\frac{\partial H_0}{\partial q_j} \frac{\partial q_j}{\partial z_j^*} + \frac{\partial H_0}{\partial p_j} \frac{\partial p_j}{\partial z_j^*} \right) \\
    &=
    -i \left(\frac{1}{\sqrt{2}}\frac{\partial H_0}{\partial q_j}  - \frac{1}{\sqrt{2} i } \frac{\partial H_0}{\partial p_j}  \right) \\
    &=
    \frac{1}{\sqrt{2}} \left( - i \frac{\partial H_0}{\partial q_j}  +  \frac{\partial H_0}{\partial q_j}  \right).
\end{align}
Noting that $\dot{q}_j = \sqrt{2} \Re[\dot{z}_j]$ and $\dot{p}_j = \sqrt{2} \Im[\dot{z}_j]$, we see that Hamilton's equations are satisfied for this term. We may write this succinctly using the symplectic form  $\Omega$ \cite{serafini_quantum_2023} as
\begin{equation}
    \dot{\bold{r}}_0 = \Omega \nabla H_0.
\end{equation}
Looking at the $H_c$ term, we find by the same process

\begin{align}
     \frac{\partial H_c}{\partial z_j^*} =
    \frac{1}{\sqrt{2}} \left(  \frac{\partial H_c}{\partial q_j}  + i \frac{\partial H_c}{\partial q_j}  \right),
\end{align}
so instead of Hamilton's equations this term gives 
\begin{equation}
    \dot{\bold{r}}_c =\nabla H_c.
\end{equation}
Therefore the classical evolution due to the Hamiltonian part for a delta function distribution is
\begin{equation}
\mathcal{L}_H^\text{(cl)}[f] = - \nabla \cdot \left(\left[\Omega \nabla H_0 +   \nabla H_c\right]f\right).
\end{equation}
Note that the first term can be written using the Poisson bracket $\{\cdot, \cdot\}_\text{P}$, as
\begin{equation}
    -\nabla \cdot ([\Omega \nabla H_0] f) = \{ H_0, f\}_\text{P}.
\end{equation}
However, when the distribution $f$ is not a delta function the last two terms of Eq. \eqref{classical-average-hamiltonian-evo} in general do not cancel. It is easily seen that Eq. \eqref{classical-average-hamiltonian-evo} is satisfied if we include the additional term $2(H_c - \braket{H_c})f$ in the generator, giving
\begin{equation}
\label{classical-NL-hamiltonian-generator}
\mathcal{L}_H^\text{(cl)}[f] = - \nabla \cdot \left(\left[\Omega \nabla H_0 +   \nabla H_c\right]f\right) + 2(H_c - \braket{H_c})f.
\end{equation}
Equation \eqref{classical-NL-hamiltonian-generator} generalizes the result from \cite{graefe_classical_2010} by including the nonlinear term corresponding to that in Eq. \eqref{full-QMES}. Finally, we generalize further by allowing for a Fokker-Planck form dissipator $\mathcal{D}^\text{(cl)}$, which plays a similar role to the Lindblad dissipator $\mathcal{D}^\text{(q)}$ in Eq. \eqref{full-QMES}

\begin{equation}
\label{full-classical-generator}
\mathcal{L}^\text{(cl)}[f] = - \nabla \cdot \left(\left[\Omega \nabla H_0 +   \nabla H_c\right]f\right) + 2(H_c - \braket{H_c})f + \mathcal{L}^\text{(cl)}[f],
\end{equation}
where
\begin{equation}
\label{classical-dissipator}
\mathcal{D}^\text{(cl)}[f] =  - \nabla \cdot (\bold{v}_\text{d}f) + \sum_{ij} \frac{\partial^2}{\partial_i \partial_j}(\Gamma_{ij}f),
\end{equation}
with drift velocity $\bold{v}_\text{d}$ and a $4 \times 4$ diffusion tensor $\Gamma$.

\section{Classical synchronization rate}
\label{classical-speed-limit-appendix}

 In the classical case, we again have
\begin{equation}
    \dot{L}^\text{(cl)} = \beta_0 \dot{E}- \dot{S}^\text{(cl)}.
\end{equation}
As in the quantum case, the first Hamiltonian term (with $H_0$) in Eq \eqref{full-classical-generator} contributes no change to the energy or the entropy, and therefore can be ignored. We first look at the effects of the second Hamiltonian term (with $H_c$). The change in energy due to this term can be expressed as
\begin{align}
    \dot{E}_c &= \int d^{2N} \bold{r} f(\bold{r}) \dot{E}_c(\bold{r}) \\
    &= 
     \int d^{2N} \bold{r} f(\bold{r}) \dot{\bold{r}}_c(\bold{r})\cdot \nabla H_0 \\
     &= \int d^{2N} \bold{r} f(\bold{r}) \nabla H_c \cdot \nabla H_0 \\
     &= \Braket{\nabla H_c \cdot \nabla H_0}.
\end{align}
It has been shown \cite{daems_entropy_1999} that in the absence of diffusion, a Liouville-type equation has entropy generation simply given by the average divergence of the drift velocity, and so
\begin{equation}
    \dot{S}^\text{(cl)}_c = \braket{\nabla^2 H_c}.
\end{equation}
Therefore the contribution of the second Hamiltonian term to $\dot{L}^{\text{(cl)}}$ is
\begin{equation}
    \dot{L}^{\text{(cl)}}_c = \beta_0 \braket{\nabla H_c \cdot \nabla H_0} -\braket{\nabla^2 H_c}.
\end{equation}
Next we consider the effects of the nonlinear term in Eq. \eqref{full-classical-generator}

\begin{align}
    \dot{L}^\text{(cl)}_\text{NL} &= 2\int d^4 \bold{r} (H_c - \braket{H_c}) f(\ln f - \ln f_{0})\\
    & = 
  2\int d^4 \bold{r} (H_c - \braket{H_c}) f\ln f - 2\int d^4 \bold{r} (H_c - \braket{H_c})f \ln f_{0}\\
  &= 
  2\int d^4 \bold{r} (H_c - \braket{H_c}) f\ln f - 2\int d^4 \bold{r} (H_c - \braket{H_c})f (-\beta_0 H_0 - \ln Z_0)\\
  & = 
  2\int d^4 \bold{r} (H_c - \braket{H_c}) f\ln f +2\ln Z_0 \left(\int d^4 \bold{r} (H_c - \braket{H_c})f\right)+2\beta_0\left( \int d^4 \bold{r} H_0(H_c - \braket{H_c}) f\right)\\
  & = 
  2\int d^4 \bold{r} (H_c - \braket{H_c}) f\ln f +2\beta_0\left( \int d^4 \bold{r} H_0(H_c - \braket{H_c}) f\right)\\
  & = 
  2\int d^4 \bold{r} (H_c - \braket{H_c}) f\ln f +2\beta_0\left( \int d^4 \bold{r} H_0 H_c f - \braket{H_c}\int d^4 \bold{r} H_0 f\right)\\
  & = 
  2\int d^4 \bold{r} (H_c - \braket{H_c}) f\ln f +2\beta_0(\braket{H_0 H_c} - \braket{H_0} \braket{H_c})\\
  & = 
  2\int d^4 \bold{r} (H_c - \braket{H_c}) f\ln f +2\beta_0\mathcal{C}_{CE}.
\end{align}
Now we have the dynamical equation for $L^\text{(cl)}$
\begin{equation}
    \dot{L}^\text{(cl)} = \beta_0 \braket{\nabla H_c \cdot \nabla H_0} -\braket{\nabla^2 H_c} + 2 \int d^4 \bold{r} (H_c - \braket{H_c}) f\ln f +2\beta_0\mathcal{C}_{CE} - \sigma_0.
\end{equation}
First, suppose that the function $H_c$ is bounded, in which case we define the infinity norm
\begin{equation}
    \|H_c\|_\infty = \max_{\bold{r}} \left| H_c(\bold{r})\right|.
\end{equation}
We then have the absolute value of the third term bounded by $4 \|H_c\|_\infty S^\text{(cl)}$, so
\begin{equation}
    \dot{L}^\text{(cl)} \leq \beta_0 \braket{\nabla H_c \cdot \nabla H_0} -\braket{\nabla^2 H_c} + 4 \| H_c\|_\infty S^\text{(cl)}_{G;E} +2\beta_0\mathcal{C}_{CE} - \sigma_0.
\end{equation}
And because $\mathcal{C}_{CE} \leq \Delta_C \Delta_E$,
\begin{equation}
    \dot{L}^\text{(cl)} \leq \beta_0 \braket{\nabla H_c \cdot \nabla H_0} -\braket{\nabla^2 H_c} + 4 \| H_c\|_\infty S^\text{(cl)}_{G;E} +2\beta_0\Delta_C \Delta_E - \sigma_0.
\end{equation}
If, on the other hand, $H_c$ is not bounded, then we may use the Cauchy Schwartz inequality to write
\begin{align}
    \left|\int d^4 \bold{r} (H_c - \braket{H_c}) f\ln f \right|
     &= \left|\int d^4 \bold{r} (H_c - \braket{H_c}) \sqrt{f}\sqrt{f}\ln f \right|\\
     &\leq \sqrt{\int d^4 \bold{r} (H_c - \braket{H_c})^2 f} \sqrt{\int d^4 \bold{r}f(\ln f)^2} \\
     & = \Delta_C \braket{\ln( f)^2},
\end{align}
so
\begin{equation}
    \dot{L}^\text{(cl)} \leq \beta_0 \braket{\nabla H_c \cdot \nabla H_0} -\braket{\nabla^2 H_c} + 2\Delta_C \braket{\ln(f)^2} +2\beta_0\mathcal{C}_{CE} - \sigma_0.
\end{equation}
Again because $\mathcal{C}_{CE} \leq \Delta_C \Delta_E$,
\begin{equation}
    \dot{L}^\text{(cl)} \leq \beta_0 \braket{\nabla H_c \cdot \nabla H_0} -\braket{\nabla^2 H_c} + 2\Delta_C \braket{\ln(f)^2} +2\beta_0\Delta_C \Delta_E - \sigma_0.
\end{equation}
\section{Quantum dimer model}
We consider the following coupling Hamiltonian
\begin{equation}
    \label{Hc-dimer-appendix}
    H_c =i \frac{k}{2} (\hat{a}^\dag_1 \hat{a}_2 +\hat{a}^\dag_2 \hat{a}_1).
\end{equation}
For Lindblad operators we use 
\begin{equation}
\hat{F}_{i-} = \hat{a}_i^2 \: \: \: \hat{F}_{i+} = (\hat{a}_i^\dag)^2. 
\end{equation}
We first find the stationary state of a single mode under these dynamics when the anti-Hermitian coupling is turned off. The single-mode dissipator is
\begin{equation}
    \mathcal{D}[\hat{\rho}] = \gamma_+\mathcal{D}_+[\hat{\rho}] + \gamma_-\mathcal{D}_-[\hat{\rho}],
\end{equation}
where
\begin{equation}
    \mathcal{D}_-[\hat{\rho}]= \hat{a}^2 \hat{\rho} (\hat{a}^\dag)^2-\frac{1}{2} \{(\hat{a}^\dag)^2 \hat{a}^2, \hat{\rho}\}
\end{equation}
and
\begin{equation}
    \mathcal{D}_+[\hat{\rho}]= (\hat{a}^\dag)^2 \hat{\rho} \hat{a}^2-\frac{1}{2} \{ \hat{a}^2 (\hat{a}^\dag)^2, \hat{\rho}\}.
\end{equation}
We compute
\begin{align}
    \mathcal{D}_-[\ket{n}\bra{n}] &= n(n-1)\ket{n-2}\bra{n-2}-n(n-1)\ket{n}\bra{n} \\
    &= n(n-1)(\ket{n-2}\bra{n-2}-\ket{n}\bra{n}),
\end{align}
and
\begin{align}
    \mathcal{D}_+[\ket{n}\bra{n}] &= (n+1)(n+2)\ket{n+2}\bra{n+2}-(n+1)(n+2)\ket{n}\bra{n} \\
    &= (n+1)(n+2)(\ket{n+2}\bra{n+2}-\ket{n}\bra{n}).
\end{align}
We assume the stationary state $\hat{\rho}_0$ is diagonal in the number basis, which ensures that the Hamiltonian term is eliminated
\begin{equation}
    \hat{\rho}_0 = \sum_{n=0}^\infty p_n \ket{n} \bra{n}.
\end{equation}
We then have
\begin{equation}
    \braket{m|\mathcal{D}_-[\hat{\rho}]|m}  = -m(m-1) p_m + (m+2)(m+1)p_{m+2}
\end{equation}
and
\begin{equation}
    \braket{m|\mathcal{D}_+[\hat{\rho}]|m}  = -(m+1)(m+2) p_m + (m-1)mp_{m-2},
\end{equation}
so
\begin{equation}
    \braket{m|\mathcal{D}[\hat{\rho}]|m} = \gamma_+ \left[m(m-1)p_{m-2}-(m+1)(m+2)p_m \right]+\gamma_-\left[(m+1)(m+2)p_{m+2}-m(m-2)p_m\right]
\end{equation}
Assuming $p_m = C \exp(-\xi m)$, we then have
\begin{align}
    \braket{m|\mathcal{D}[\hat{\rho}]|m} &= \gamma_+ m(m-1)\exp(2\xi) p_m  - \left[ \gamma_+ (m+1)(m+2) + \gamma_- m(m-1)\right]p_m + \gamma_- (m+1)(m+2)\exp(-2\xi)\\
    &=\left[\gamma_+\exp(2\xi) - \gamma_+ - \gamma_- + \gamma_- \exp(-2\xi)\right]p_m m^2 + \left[ -\gamma_+ \exp(2\xi) - 3 \gamma_+ + \gamma_- + 3\gamma_- \exp(-2\xi)\right]p_m m \\ &+ \left[ -2\gamma_+ 
    + 2\gamma_- \exp(-2\xi)\right].
\end{align}
We set $\braket{m|\mathcal{D}[\hat{\rho}]|m} = 0$ for all $m$, and therefore the terms of order zero, one, and two in $m$ must be identically zero. We find
\begin{equation}
    2 \gamma_- \exp(-2\xi) = 2 \gamma_+
\end{equation}
\begin{equation}
    \xi = \frac{1}{2} \ln \left( \frac{\gamma_-}{\gamma_+}\right).
\end{equation}
This is simply the Gibbs state of the harmonic oscillator with inverse temperature
\begin{equation}
    \beta = \frac{\xi}{\omega} = \frac{1}{2\omega}\ln \left(\frac{\gamma_-}{\gamma_+} \right).
\end{equation}
Because there is no interaction by assumption, the Gibbs state of the combined system is the tensor product
\begin{equation}
    \hat{\rho}_0 = \bigotimes_{j=1}^N \hat{\rho}_{G;\beta_j},
\end{equation}
where
\begin{equation}
    \beta_j = \frac{1}{2\omega_j} \ln \left( \frac{\gamma_{j-}}{\gamma_{j+}}\right).
\end{equation}
If we fix $\gamma_{j+} = \gamma_+$ and treat the frequencies $\omega_j$ as free parameters, may then choose
\begin{equation}
    \gamma_{j-} = \exp(2 \beta \omega_j) \gamma_+,
\end{equation}
which is the form of the fluctuation dissipation theorem for this system. The commutator which appears in Eq. \eqref{QSL-unbounded} is

\begin{align}
    [H_0, H_c] &= i \frac{k}{2} [\omega_1\hat{a}^\dag_1 \hat{a}_1 + \omega_2\hat{a}^\dag_2 \hat{a}_2,\hat{a}^\dag_1 \hat{a}_2 +\hat{a}^\dag_2 \hat{a}_1] \\
    & = i \frac{k}{2}(\omega_1a_1^\dag a_2+\omega_2 a_2^\dag a_1 - \omega_1 a_2^\dag a_1 - \omega_2 a_1^\dag a_2)\\
    &= i \frac{k}{2}(\omega_1 - \omega_2)(a_1^\dag a_2-a_2^\dag a_1),
\end{align}
so
\begin{equation}
   \dot{L} \leq 2 \Delta_C \sqrt{\mathcal{E}+S_{G;E}^2}+2\beta_0\sqrt{ \Delta_E^2 \Delta_C^2 -  \frac{1}{8}k^2|\omega_1-\omega_2|^2 \left|\braket{a_1^\dag a_2-a_2^\dag a_1}\right|^2} - \sigma_0.
\end{equation}

\begin{figure*}
    \centering
    \includegraphics[width=17.2cm]{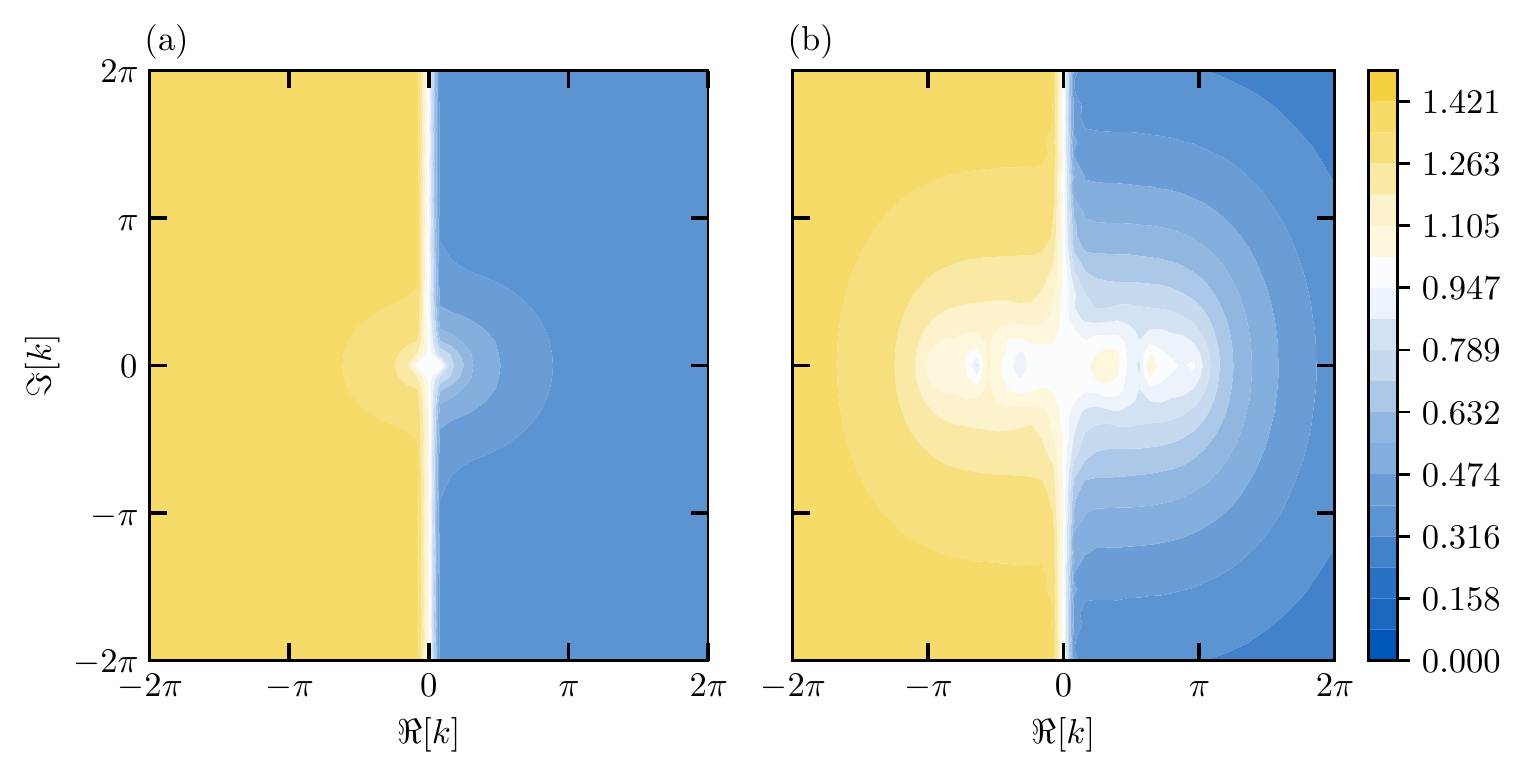}
    \caption{(Color online) Distance measure $D$ at time $t = 10$ for quantum (a) and classical (b) evolution, as a function of the real and imaginary part of $k$. Frequencies are $\omega_1 = \pi$, $\omega_2 = 2\pi$. Classical synchronous regime is $\Re[k]|>\pi$ for $\Im[k]=0$. Quantum system synchronizes for smaller $\Re[k]$.}
    \label{fig:random-chi-DS}
\end{figure*}

\section{Classical dimer model}
In the classical limit of the dimer model, the function  $H_c$ is
\begin{equation}
    H_c = \frac{k}{2}(x_1 x_2 + p_1 p_2),
\end{equation}
and
\begin{equation}
    H_0 = \frac{\omega_1}{2}(x_1^2 + p_2^2) + \frac{\omega_2}{2}(x_2^2+p_2^2).
\end{equation}
Therefore
\begin{equation}
    \nabla H_0 = (\omega_1 x_1, \omega_1 p_1, \omega_2 x_2, \omega_2 p_2)^T,
\end{equation}
while
\begin{equation}
    \nabla H_c = \left(\frac{k}{2} x_2, \frac{k}{2} p_2, \frac{k}{2} x_1, \frac{k}{2} p_1\right)^T,
\end{equation}
We then find
\begin{equation}
    \nabla H_0 \cdot \nabla H_c = \frac{k}{2}(\omega_1 x_1 x_2 + \omega_2 p_2 p_2 + \omega_2 x_1 x_2 + \omega_2 p_1 p_2) = \frac{k(\omega_1 + \omega_2)}{2}\bold{r}_1 \cdot \bold{r}_2,
\end{equation}
and
\begin{equation}
    \nabla^2 H_c = 0,
\end{equation}
so

\begin{equation}
    \dot{L}^\text{(cl)} \leq \beta_0 \frac{k(\omega_1+\omega_2)}{2}\braket{\bold{r}_1 \cdot \bold{r}_2}+ 2\Delta_C \sqrt{\braket{\ln(f)^2} }+2\beta_0\mathcal{C}_{CE} - \sigma_0.
\end{equation}

\end{document}